\documentclass[journal=jctcce,manuscript=article,layout=twocolumn]{achemso}
\usepackage{xcolor}
\usepackage{graphicx}
\usepackage{float}
\usepackage{mwe}
\usepackage{soul}
\usepackage{indentfirst}
\usepackage{algorithm}
\usepackage{algpseudocode}
\usepackage{amsfonts, amsmath, amssymb}
\usepackage[raggedrightboxes]{ragged2e}
\usepackage[normalem]{ulem}
\usepackage{threeparttable}
\usepackage{tabularray}
\usepackage{xurl}
\usepackage{adjustbox}
\usepackage{multirow}
\usepackage{tikz}
\usetikzlibrary{shapes,arrows}
\usepackage[version=3]{mhchem}
\usepackage{amsmath}
\usepackage{graphicx}
\usepackage{multirow}
\usepackage[english]{babel}
\usepackage[T1]{fontenc}
\usepackage[utf8]{inputenc}

\usepackage{hyperref}
\hypersetup{
 bookmarks=true,		
 unicode=false,			
 pdftoolbar=true,		
 pdffitwindow=false,		
 pdfstartview={FitH},		
 pdfcreator={pdflatex},		
 pdfproducer={vim},		
 pdfnewwindow=true,		
 colorlinks=true,		
 linktoc=page,			
 linkcolor=blue,		
 citecolor=blue,		
 filecolor=blue,		
 urlcolor=blue			
}

\author{\'Ad\'am Ganyecz}
\affiliation{%
Strongly Correlated Systems "Lendület" Research Group,
Wigner Research Centre for Physics, H-1525, Budapest, Hungary
}%
\email{ganyecz.adam@wigner.hun-ren.hu}
\author{Zsolt Benedek}
\affiliation{Department of Physics of Complex Systems, Eötvös Loránd University, Egyetem tér 1-3, H-1053 Budapest, Hungary}
\alsoaffiliation{MTA–ELTE Lend\"{u}let "Momentum" NewQubit Research Group, Pázmány Péter, Sétány 1/A, 1117 Budapest, Hungary}
\author{Kl\'ara Petrov}
\affiliation{%
Department of Physical Chemistry and Materials Science, Faculty of Chemical Technology and Biotechnology, Budapest University of Technology and Economics, Műegyetem rkp. 3., H-1111 Budapest, Hungary
}%
\author{Gergely Barcza}
\affiliation{%
Strongly Correlated Systems "Lendület" Research Group,
Wigner Research Centre for Physics, H-1525, Budapest, Hungary
}%
\author{Andr\'as Olasz}
\affiliation{%
Strongly Correlated Systems "Lendület" Research Group,
Wigner Research Centre for Physics, H-1525, Budapest, Hungary
}%
\alsoaffiliation{Furukawa Electric Institute of Technology Ltd., Késmárk street 28/A,
H-1158 Budapest, Hungary
}
\author{Mikl\'os A. Werner}
\affiliation{%
Strongly Correlated Systems "Lendület" Research Group,
Wigner Research Centre for Physics, H-1525, Budapest, Hungary
}%
\author{Örs Legeza}
\affiliation{%
Strongly Correlated Systems "Lendület" Research Group,
Wigner Research Centre for Physics, H-1525, Budapest, Hungary
}%
\alsoaffiliation{Dynaflex Ltd., H-1028 Budapest, Hungary
}
\alsoaffiliation{
Institute for Advanced Study,Technical University of Munich, Lichtenbergstrasse 2a, 85748 Garching, Germany
}
\alsoaffiliation{Parmenides Stiftung, Hindenburgstr. 15, 82343, Pöcking, Germany
}

\title[]{
Assessing the Reliability of Truncated Coupled Cluster Wavefunction: Estimating the Distance from the Exact Solution}

\begin{document}

\begin{abstract}
A new approach is proposed to assess the reliability of the truncated wavefunction methods by estimating the deviation from the full configuration interaction (FCI) wavefunction. While typical multireference diagnostics compare some derived property of the solution with the ideal picture of a single determinant, we try to answer a more practical question, how far is the solution from the exact one. Using the density matrix renormalization group (DMRG) method to provide an approximate FCI solution for the self-consistently determined relevant active space, we compare the low-level CI expansions and one-body reduced density matrixes to determine the distance of the two solutions ($\Tilde{d}_\Phi$, $\Tilde{d}_\gamma$). We demonstrate the applicability of the approach for the CCSD method by benchmarking on the W4-17  dataset, as well as on transition metal-containing species.
We also show that the presented moderate-cost, purely wavefunction-based metric is truly unique in the sense that it does not correlate with any popular multireference measures.
We also explored the usage of CCSD natural orbitals ($\Tilde{d}_{\gamma,\mathrm{NO}}$) and its effect on the active space size and the metric.
The proposed diagnostic can also be applied to other wavefunction approximations, and it has the potential to provide a quality measure for post-Hartree-Fock procedures in general.

\end{abstract}

\section{Introduction}
\label{sec:Introduction}

In computational quantum chemistry, correlation effects are often distinguished into dynamic and static correlation according to their character~\cite{Mok-1996}.
The latter one, also known as nondynamical correlation, can be further divided to Type A or left-right strong correlation and Type B or angular strong correlation \cite{Scuseria-2009,Hollett-2011}.
Type A correlation occurs when there is an absolute near-degeneracy, for example, bond stretching, while Type B refers to relative near-degeneracy when the gap is small compared to the orbital energies.

When dynamic correlation dominates, the routinely applied single reference (SR) quantum chemical methods, such as density functional theory (DFT) or the gold standard coupled cluster singles doubles and perturbative triples (CCSD(T)), work well.
However, they often do not provide accurate results for molecules with a significant nondynamical correlation.~\cite{Margraf-2017,Cremer-2002} Therefore, measuring or at least estimating the degree of correlation is a crucial point in choosing the appropriate methodology. From the practical point of view in computational chemistry, it is also required that the cost of such a measurement should be comparable to that of a single reference calculation.~\cite{Duan-2020}

To address this issue, numerous multireference (MR) diagnostics appeared in the literature \cite{Jiang-2012,Fogueri-2013,Sprague-2014} that attempt to predict the validity of the single-reference approach by analyzing the SR results themselves.
Only two of them are based on CI coefficients ($c_i$): $C_0^2$\cite{Sears-2008a,Sears-2008b} is the weight of the leading determinant, usually from a CASSCF or CASCI calculation, while MR\cite{Coe-2014} uses the second and fourth power of $c_i$ coefficients to determine the deviation from the single determinant wavefunction.

Specifically, for CC methods, the most straightforward metrics were the maximum of $t_1$ singles and $t_2$ doubles amplitudes, which were available in the early versions of ACES quantum chemical program suite\cite{Stanton-1992,Perera-2020}.
The well-known $T_1$\cite{Lee-1989} and $D_1$\cite{Janssen-1998,Leininger-2000,Lee-2003} are based on the Euclidean and the Frobenius norm of $t_1$ singles organized into the appropriate vector and matrix, respectively.
Double excitation amplitudes can be used in a similar manner to obtain $T_2$\cite{Coe-2015} and $D_2$\cite{Nielsen-1999} for closed-shell systems.
Most recently, $S$ diagnostic was proposed, which also uses Lagrange multipliers in addition to the cluster amplitudes\cite{Faulstich-2023}.

Another class of metrics is based on occupation numbers, usually of natural orbitals, which can be obtained from various methods. The simplest example uses the occupation number of HOMO and LUMO\cite{Tishchenko-2008,Fogueri-2013,Jensen-1988}, which can be combined in one number, $M$\cite{Tishchenko-2008}.
Coe and Paterson (based on Löwdin) defined $\Theta$, which indicates how far the wavefunction is from the single-determinant solution of Hartree-Fock\cite{Lowdin-1955,Coe-2015}.
Ramos-Cordoba et al. derived $I_\mathrm{nd}$ and $I_\mathrm{d}$
nondynamical and dynamical correlation indices based on the two-particle cumulant matrix\cite{Ramos-Cordoba-2016,Ramos-2017}.
Kesharwani and co-workers defined an intensive quantity, $r_\mathrm{nd}$, as a ratio of nondynamical and total correlation indices, $I_\mathrm{nd}/(I_\mathrm{nd}+I_\mathrm{d})$.
Grimme and Hansen \cite{Grimme-2015,Bauer-2017} proposed a function called fractional occupation number weighted density (FOD) to visualize static or nondynamical correlation in the molecule. Its spatial integration produces a single number that can be used as an MR diagnostic metric. The fractional occupational numbers are determined with finite-temperature DFT in this approach.
Bartlett et al.\cite{Bartlett-2020} defined several new metrics for CC based on natural orbital occupation numbers, determined from 1-body RDM, but these definitions can also be used with any occupation number.
The external electron number (EEN) shows how many electrons are in the virtual orbitals, its counterpart, the variance relative to the ideal single determinant case, $\hat{V}$, is based on the occupation numbers of occupied orbitals and scaled with the number of orbitals. NON is the largest occupation number on unoccupied natural orbitals, which is also closely related to the LUMO occupation number. They also constructed a multireference index (MRI), transforming the occupation numbers to a single number ranging from -1 to 1, where MRI around -1 suggests MR character, while around +1 SR character is expected.

Following the work of Legeza and S\'olyom on single orbital entropies~\cite{Legeza-2003b} and their sum providing total correlation~\cite{Legeza-2004b}
Boguslawski and coworker used these concepts to study multireference nature of strongly correlated molecules~\cite{Boguslawski-2013}. Later, Stein and Reiher used total correlation to define $Z_{s(1)}$ diagnostic which scales so that the maximal entanglement refers to 1 while no entanglement refers to the value 0~\cite{Stein-2017}.

The next group of metrics can be called energy-based multireference diagnostics. These use some ratio of total atomization energies.
\%TAE(T) \cite{Karton-2006,Karton-2011} measures the effect of perturbative triples on total atomization energy.
$B_1$\cite{Schultz-2005} uses the energy difference of BLYP\cite{Becke-1988,Lee-1988} and B1LYP\cite{Adamo-1997} and scales with the number of bonds.
$A_\lambda$\cite{Fogueri-2013} takes the difference  between the pure and its hybrid DFT version and is scaled with $\lambda$, which is a percent of the HF-exchange of the hybrid functional.
Recently, Martin and his co-workers proposed \%TAE$_X$~\cite{Martin-2022}, which measures for a given HF density the difference of the DFT and HF exchange energies.
The various multireference diagnostics used in this work are summarized in Table \ref{tab:MR_diagnostics}.

Even though numerous diagnostics exist, their accuracy in predicting which system should not be handled with single reference methods is often questionable. Usually, they do not even correlate with each other\cite{Fogueri-2013,Duan-2020,Martin-2022}. This leads to arbitrary rules where multiple diagnostics are used, such as species with 3d transition metals are considered multireference if $T_1$ > 0.05, $D_1$ > 0.15, and\%TAE[(T)] > 10\%.\cite{Jiang-2012}

All of the aforementioned metrics try to determine the degree of multireference character by taking the ideal single reference wavefunction and measuring the deviation from it. However, the real question is how far the solution of a given method is from the exact full CI wavefunction.

In this work, we propose a new family of metrics, $\Tilde{d}$, which represents the quality of the wavefunction by measuring its deviation from the reference obtained by high-level theory. $\Tilde{d}_\Phi$ uses CI-coefficients up to double excitations, while $\Tilde{d}_\gamma$ uses the 1-body reduced density matrix, and $\Tilde{d}_{\gamma,\mathrm{NO}}$ uses also the natural orbital transformation to reduce the size of the active space used. First, we present the formulation of the new metrics. Then, we show the validity of the introduced approximations. After that, we compare the performance of $\Tilde{d}$-s to other popular multireference diagnostics on the W4-17 dataset, and also a grouping of existing metrics is presented. The performance is also demonstrated on transition metal species, which are particularly difficult cases for standard single reference methods.

\begin{table*}

    \centering
    \centering
    \caption{List of MR diagnostics studied in this work. }
    \scalebox{0.85}{
    \begin{threeparttable}
    \begin{tabular}{p{2cm}|p{4cm}p{8cm}p{5cm}l}
    Name  & Type & Definition\tnote{a}  & This work\tnote{b}   & Ref   \\
    \hline
     $C_0^2$  & CI coefficient & square of coefficient of leading determinant & CCSD, DMRG    &  \citenum{Sears-2008a,Sears-2008b}   \\
    $MR$ & CI coefficient & $\sum\limits_i |c_i|^2-|c_i|^4$  & CCSD, DMRG &  \citenum{Coe-2014}\\
     max $|t_1|$ & CC amplitudes & maximum of $t_1$ amplitudes  & CCSD & \citenum{Stanton-1992} \\
     max $|t_2|$ & CC amplitudes & maximum of $t_2$ amplitudes  & CCSD &  \citenum{Stanton-1992} \\
     $T_1$ & CC amplitudes & $||t_1||_F/\sqrt{n_\mathrm{corr}}$ & CCSD & \citenum{Lee-1989} \\
     $T_2$ & CC amplitudes & $||t_2||_F/\sqrt{n_\mathrm{corr}}$ & CCSD & \citenum{Coe-2015} \\
     $D_1$ & CC amplitudes & $||\mathbf{T}_1||_2$ & CCSD & \citenum{Janssen-1998,Leininger-2000,Lee-2003} \\
     $D_2$ & CC amplitudes & $\mathrm{max}(||\mathbf{T}_2^o||_2,||\mathbf{T}_2^v||_2)$, \newline where $\mathbf{T}_2^o\in \mathbb{R}^{ov^2 \times o}$  and $\mathbf{T}_2^v\in \mathbb{R}^{o^2v \times v}$& CCSD & \citenum{Nielsen-1999} \\
     $n_{\mathrm{HOMO}}$ & occupation numbers & occupation number of HOMO &  CCSD, DMRG, FT-DFT & \citenum{Tishchenko-2008,Fogueri-2013,Jensen-1988}  \\
    $n_{\mathrm{LUMO}}$ & occupation numbers & occupation number of LUMO &  CCSD, DMRG, FT-DFT & \citenum{Tishchenko-2008,Fogueri-2013,Jensen-1988}  \\
     $M$ & occupation numbers & $\frac{1}{2}(2-n_{\mathrm{HOMO}}+n_{\mathrm{LUMO}}+\sum\limits_{j}^{N_{\mathrm{SOMO}}}|n_j-1|)$ & CCSD, DMRG, FT-DFT & \citenum{Tishchenko-2008} \\
    $I_{\mathrm{nd}}$ & occupation numbers & $\frac{1}{2}\sum\limits_{\sigma,i}n_i^\sigma(1-n_i^\sigma)$ & CCSD, DMRG, FT-DFT & \citenum{Ramos-Cordoba-2016,Ramos-2017} \\
     $r_{\mathrm{nd}}$ & occupation numbers & $ \frac{\frac{1}{2}\sum\limits_{\sigma,i}n_i^\sigma(1-n_i^\sigma)}{\frac{1}{4}\sum\limits_{\sigma,i}[n_i^\sigma(1-n_i^\sigma)]^{1/2}}$  & CCSD, DMRG, FT-DFT & \citenum{Kesharwani-2018} \\
     $\Theta$ & occupation numbers & $1-\frac{1}{n} \sum\limits_{i,\sigma}n_i^{\sigma2}$& CCSD, DMRG, FT-DFT& \citenum{Lowdin-1955,Coe-2015} \\
    $N_\mathrm{FOD}$ & occupation numbers & $\sum\limits_i^{N_{\mathrm{occ}}} 1-n_i^\sigma+\sum\limits_j^{N_{\mathrm{virt}}} n_j^\sigma $ & CCSD, DMRG, FT-DFT & \citenum{Grimme-2015,Bauer-2017}  \\
EEN & occupation numbers & $\sum\limits_i^{N_{virt}} n_i^\sigma$ & CCSD, DMRG, FT-DFT & \citenum{Bartlett-2020}   \\
$\hat{V}$ & occupation numbers & $\frac{1}{n} (\sum\limits_i^{N_{occ}}n_i^\sigma-\sum\limits_i^{N_{occ}}n_i^{\sigma2})$ & CCSD, DMRG, FT-DFT & \citenum{Bartlett-2020}   \\
MRI & occupation numbers & ${\rm MRI}=-\tanh (4.016+\log(I))\newline  I=\sum\limits_i\exp(-1500\times(0.5-n_i)^6)$ & CCSD, DMRG, FT-DFT & \citenum{Bartlett-2020}  \\
NON & occupation numbers &largest occupation number on unoccupied natural orbitals & CCSD, DMRG, FT-DFT & \citenum{Bartlett-2020}   \\

     $Z_{s(1)}$ & orbital entropy & $\frac{1}{N \mathrm{ln} 4}\sum\limits_i^{N} s_i(1)$ & CCSD, DMRG& \citenum{Stein-2017} \\

     \%TAE(T) & energy & $100\times\frac{\mathrm{TAE[CCSD(T)]}-\mathrm{TAE[CCSD]}}{\mathrm{TAE[CCSD]}}$ & CCSD(T) & \citenum{Karton-2006,Karton-2011} \\
     $B_1$ & energy & $\frac{\mathrm{TAE[BLYP]-TAE[B1LYP]}}{n_{\mathrm{bonds}}}$ &  BLYP, B1LYP &  \citenum{Schultz-2005}  \\
     $A_\lambda$ & energy & $\frac{100}{\lambda}\frac{\mathrm{TAE[XC]-TAE[X_\lambda C]}}{\mathrm{TAE[XC]}}$ & PBE, PBE0 & \citenum{Fogueri-2013} \\

    \end{tabular}
    \begin{tablenotes}
    \item[a]
    $N$ ($N_\mathrm{occ}$, $N_\mathrm{virt}$, $N_\mathrm{SOMO}$): number of orbitals (occupied, virtual, single occupied), \\
    $n_{(corr)}$: number of (correlated) electrons,
    $n_i$ ($n_i^\sigma$ $n_\mathrm{HOMO}$, $n_\mathrm{LUMO}$) : occupational number (spinorbitals, HOMO, LUMO) \\
    $c_i$: CI coefficient,
    $t_1$, $t_2$: single/double excitation CC amplitudes,
    TAE: total atomization energy
    \item[b] Calculations used in this work to determine the metric
    \end{tablenotes}

 \end{threeparttable}}
 \label{tab:MR_diagnostics}
\end{table*}

\section{Theory}
\label{sec:Theory}

In this section, we present how the new metrics are formulated. First, we discuss the reference used, and then which property will be used as a basis of comparison. After that we will discuss the orbital selection procedure and then present the definition of the final metric.

\subsection{Reference}

Testing the quality of approximate quantum chemical methods, such as coupled cluster theory, against the exact reference provided by FCI is possible up to only 20 orbitals due to its computational demand~\cite{Vogiatzis-2017}.
To overcome this limitation, in this paper, we applied the density matrix renormalization group (DMRG) approach~\cite{White-1999} as a robust quasi-FCI solver with polynomial scaling.
The reference for all the investigated quantities was derived from the high-precision DMRG wavefunction whose accuracy was controlled by the truncation error~\cite{Legeza-2003}.
For a practical review of the DMRG method, see Refs.~\citenum{Legeza-2008,Chan-2008,Yanai-2009,Marti-2010c,Wouters-2014,Szalay-2015a,Olivares-2015,Baiardi-2020,Cheng-2022}.

\subsection{Basis of comparison}

There are numerous ways to compare wavefunctions. The easiest way would be to use energy or some other property. However, in this case, we would lose most of the information that is contained in the wavefunction. If the wavefunction is expressed in MO-based Slater-determinants, then one can use the CI-coefficients ($c_i$) as a basis of comparison.
Due to the exponential formulation of CC wavefunctions, studied in our work in detail, contributions for all higher excited determinants are present even in for CC truncated to single and double (SD) excitations. The expression of the full $c_i$ is, however, numerically infeasible for the systems of our interest, as it would be as large as the FCI wavefunction, therefore it should also be truncated. Additionally, due to the internal normalization of the CC wavefunction, we can not determine the error caused by this truncation, unless we evaluate all coefficients. Consequently, the comparison based on the CI coefficients will be restricted up to double excitations by projecting the wavefunctions to this subspace and then normalizing them,
\begin{equation}
    |\Phi\rangle=\frac{\hat{P}|\Psi\rangle}{||\hat{P}|\Psi\rangle||_2} \; .
\end{equation}
where $|\Psi\rangle$ is the wavefunction expressed as the linear combination of all possible configurations. In the following, $|\Phi\rangle$ will denote the wavefunction projected on the basis of the reference determinant and the corresponding single and double excited configurations and then normalized, $\hat{P}$ projects to the subspace of single and double excitations, and $||\phantom{a}||_2$ stands for the usual norm of the wavefunction.

Reduced density matrices provide an alternative way to represent a quantum state. The one-particle reduced density matrix (1-RDM, $\gamma$) contains partial, compressed information about the wavefunction, and will also be used as an alternative basis of comparison. As CC is not a variational method, different formulations exist for the density matrices. In this work we used unrelaxed RDMs, i.e. response RDMs without orbital relaxaton effects.

As a distance metric, we will use the Euclidean norm of the differences of the normalized vector made of $c$ coefficients ($\Psi$), or their projection $c^\mathrm{SD}$ up to doubles excitation ($\Phi$), or the Hilbert-Schmidt norm of the differences of 1-body RDM matrices ($\gamma$). The distance from the reference wavefunction or RDM is denoted by $d$ while the basis of comparison is represented in the lower index.

\begin{subequations}
    \begin{align}
        d_\Psi &=||\Psi_\mathrm{CCSD}-\Psi_\mathrm{DMRG}|| \\
        d_\Phi &=||\Phi_\mathrm{CCSD}-\Phi_\mathrm{DMRG}|| \\
        d_\gamma &=||\gamma_\mathrm{CCSD}-\gamma_\mathrm{DMRG}||
    \end{align}
\end{subequations}

We note that $d_\Psi$ is only available for small systems, but $d_\Phi$ and $d_\gamma$ can be evaluated efficiently for larger systems too. In the following, if there is no subscript following $d$, we refer to the distance metric in general.

\subsection{Active space selection}

Although DMRG can handle active spaces close to hundred orbitals~\cite{Wouters-2014,Li-2019,Brabec-2021,Baiardi-2020,Menczer-2024b,Menczer-2024x,Legeza-2025}, the accessible number of orbitals is still lower than in CCSD, therefore the orbital space should be truncated.

A universally applicable black-box active space selection strategy is challenging to design.
However, an automatized selection protocol based on single orbital entropies~\cite{Legeza-2003b} introduced via the dynamically extended active space (DEAS) procedure, also utilized in a more general framework~\cite{Stein-2017,Faulstich-2019}, can offer a reasonable solution.
In this procedure, an initial, low bond dimension DMRG is performed on the full valence space, and the active space is selected based on single-orbital entropies with an empirical pre-set
cutoff relative to the highest observed entropy value.

In the present study, we develop an alternative entropy-based selection. As a first step of the entire workflow, we perform a standard CCSD (or any other post-HF) calculation, which we later want to analyze and from which the $\Phi$ or $\gamma$ can be easily extracted (light blue part of Fig. \ref{fig:flowchart}).
Recall that the single-orbital entropy
~\cite{Legeza-2003b}
for orbital $i$ is defined as
\begin{equation}
    s_\mathrm{i}(1)=-\sum_\mathrm{\alpha} p_i^\alpha \mathrm{ln} p_i^\alpha,
\end{equation}
where $p_i^\alpha$ denotes the probability that orbital $i$ is found in occupation state $\alpha\in \{0, \downarrow, \uparrow, \downarrow \uparrow\}$ in the many-body wavefunction. The $p_i^\alpha$ weights can be determined by summing the square of CI coefficients of the corresponding determinants. We note that in practical CC calculations, we estimate the probabilities by restricting the CI expansion of the wavefunction up to double excitations. Alternatively, exact entropies could also be constructed from relevant entries of the 1 and 2-body reduced density matrices\cite{barcza2015entanglement,Boguslawski-2015,Nowak-2021}; however, this requires solving the $\Lambda$ equations  besides the CC amplitude equations, which both have similar costs, while for the calculation of the energy the solution of the CC amplitude equations are sufficient.

By sorting the $s_{i}(1)$ values in descending order, the orbitals can be selected according to their chemical relevance, where orbitals with the highest entropies will form the active space.

However, the question remains how to choose the cutoff of $s_{i}(1)$ between active and inactive orbitals. Here, instead of predefining a fixed $s_{i}(1)$ value or proportion - such as
for example 10\% in the AutoCAS program \cite{Stein-2017} -, we apply a different approach.
Based on the sorted $s_{i}(1)$ values, we identify various subsets of orbitals by their importance, where active orbitals are distinguished from inactive ones by the rate of the change in the entropy profile.
More precisely,  orbitals labeled $\{1,\ldots, i\}$  in the descending entropy order define an active space in case of $s_{i-2}(1)-s_{i-1}(1)<s_{i-1}(1)-s_{i}(1)$ (i.e., the entropy difference between two adjacent orbitals is larger than previously).

In some way, we try to separate the static and dynamic correlation. In that sense, we are interested in how well the static correlation is described by the selected active space, and not in the dynamic correlation mostly captured by the inactive orbitals.
Once the possible active spaces are determined, starting from the smallest ones, we perform CCSD calculation for the selected active space, which results are marked by superscript AS. The obtained CCSD wavefunction ($\Phi^\mathrm{AS}_{\rm CCSD}$) or RDM ($\gamma^\mathrm{AS}_\mathrm{CCSD}$) is compared to the corresponding projection of the full space CCSD wavefunction ($\Phi^\mathrm{FS\rightarrow AS}_{\rm CCSD}$ or $\gamma^\mathrm{FS\rightarrow AS}_\mathrm{CCSD}$). The corresponding distances are
\begin{subequations}
\begin{align}
d^\mathrm{AS}_{\Phi} &=||\Phi^\mathrm{AS}_{\rm CCSD}-\Phi^\mathrm{FS \rightarrow AS}_{\rm CCSD}|| \; , \\
d^\mathrm{AS}_\gamma &=||\gamma^\mathrm{AS}_{\rm CCSD}-\gamma^\mathrm{FS \rightarrow AS}_{\rm CCSD}|| \; .
\end{align} \label{eq:d_AS}
\end{subequations}
Note, that the full space $\Phi_\mathrm{CCSD}$ and $\gamma_\mathrm{CCSD}$ are projected to the selected subspace, where only those configurations are kept, in which excitations are allowed for orbitals in the active space, hence the "$\mathrm{FS}\rightarrow\mathrm{AS}$" in the upper index. The wavefunction and RDM are normalized to give 1 and the correct electron number, respectively.

By comparing wavefunctions of the full space CCSD and the one of the CCSD on active space, $d^\mathrm{AS}$ measures the error of active space selection. By definition, $d^\mathrm{AS}$ equals zero if all orbitals are active. Hence, the active space can be considered as converged once $d^\mathrm{AS}$ falls below a predefined threshold. If $d^\mathrm{AS}$ is larger than the desired threshold, we continue with including the next batch of orbitals in the active space proposed by the entropy profile, until the defined criteria are satisfied. This selection approach ensures that we select a subspace that correctly represents the whole space. For small active spaces, it might happen that first $d^\mathrm{AS}$ increases (see later), therefore we apply a conservative approach, and $d^\mathrm{AS}$ have to be not only below a threshold, but have to be decreasing (be smaller than the previous iteration). The whole active space selection is summarized by the light pink part of Fig. \ref{fig:flowchart}.

\begin{figure}
    \centering
    \includegraphics[width=1.0\linewidth]{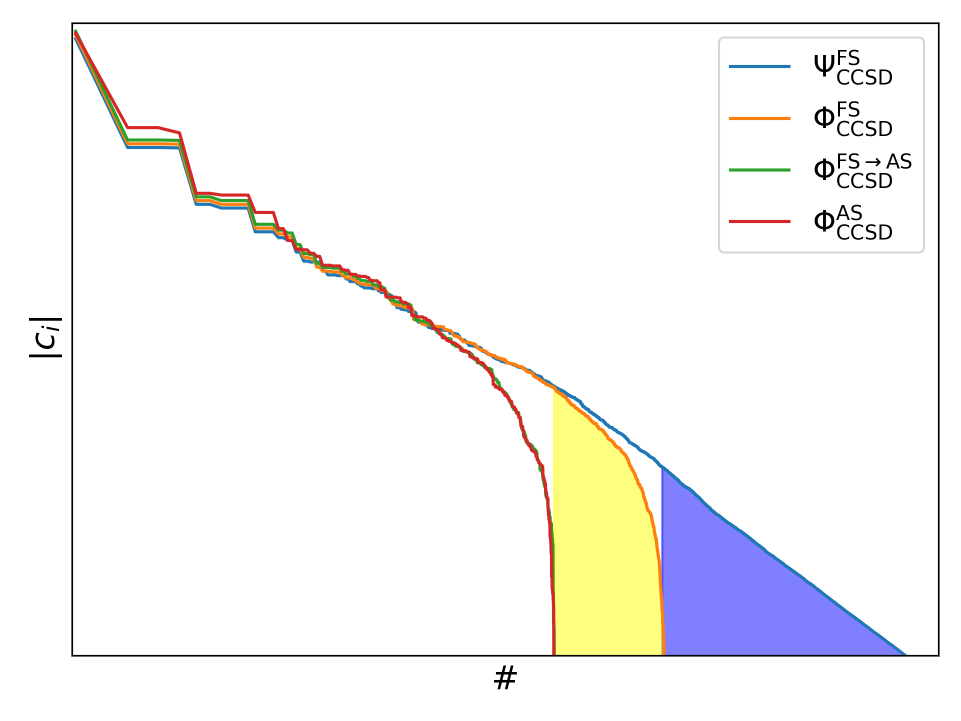}
    \caption{Schematic representation of the different normalized CCSD wavefunctions by plotting the corresponding CI coefficients for BN at CCSD/cc-pVDZ level of theory and AS represents 14 orbitals. Blue shaded part represents the part of the wavefunction which is discarded by restricting up to double excitations. Yellow shaded part shows the determinants which are discarded due to the active space truncation.}
    \label{fig:ccsd}
\end{figure}

The relation of the various wavefunctions are summarized on Fig. \ref{fig:ccsd}, using BN at CCSD/cc-pVDZ level of theory as an example. First, we have $\Psi^\mathrm{FS}_\mathrm{CCSD}$ the full space CCSD wavefunction (accessible only for small systems). Though the number of cluster amplitudes are limited, due to the exponential ansatz the CI coefficients are not. The evaluation of all coefficients is not practical, and it is limited up to double excitations; higher excitations, represented in blue shade, are ignored, and the error introduced with this truncation is unknown. $\Phi^\mathrm{FS}_\mathrm{CCSD}$ is the wavefunction with only single and double excitations, while $\Phi^\mathrm{FS\rightarrow AS}_\mathrm{CCSD}$ contains only the excitations that are in the active space. The error of this truncation is the yellow shaded area in the figure. Finally, we compare this wavefunction to $\Phi^\mathrm{AS}_\mathrm{CCSD}$, to get $d^\mathrm{AS}_\Phi$. We note that in the case of $\gamma$, there is only active space truncation and no exclusion of higher excitations.

\subsection{Forming the final metric}

Following the active space selection, the DMRG calculation is performed.
This produces a third wavefunction, besides the existing active space and full space CCSD wavefunctions, which wavefunction we represent by its CI coefficients, $c$, or 1-RDM, $\gamma$. To ensure comparability, both three wavefunctions are vectors of the same active space, are normalized to one and the electron number in 1-RDM's is also properly set.

Between the three wavefunction three distances can be defined, one of them, $d^{AS}$, is already defined in Eq.~\eqref{eq:d_AS}.
It is easy to see, that as the size of the active space approaches full space the wavefunction of the two CCSD calculation become more and more similar,
\begin{subequations}
\begin{align}
        \lim_{\mathrm{AS}\rightarrow\mathrm{FS}}  \Phi^\mathrm{AS}_\mathrm{CCSD}=\Phi^\mathrm{FS}_\mathrm{CCSD} \; , \\
        \lim_{\mathrm{AS}\rightarrow\mathrm{FS}}  \gamma^\mathrm{AS}_\mathrm{CCSD}=\gamma^\mathrm{FS}_\mathrm{CCSD} \; ,
\end{align}
\end{subequations}
and, consequently,
\begin{subequations}
\begin{align}
        \lim_{\mathrm{AS}\rightarrow\mathrm{FS}}  d^\mathrm{AS}_\Phi=0 \; ,\\
        \lim_{\mathrm{AS}\rightarrow\mathrm{FS}}  d^\mathrm{AS}_\gamma=0 \; .
\end{align}
\end{subequations}
It is important to note that the active space selection is based on the orbital entropies, meaning orbitals are included based on their importance in the wavefunction. The goal is to include all orbitals which are important to describe static correlations, while we aim to discard orbitals which are only contributing to the so-called dynamical correlation.

The other two are the distances from the DMRG solution,
\begin{subequations}
\begin{align}
    d^\mathrm{a}_\Phi&=||\Phi_\mathrm{CCSD}^\mathrm{FS \rightarrow AS}-\Phi_\mathrm{DMRG}^\mathrm{AS}|| \; \label{eq:dphi_a} ,\\
   d^\mathrm{a}_\gamma&=||\gamma_\mathrm{CCSD}^\mathrm{FS \rightarrow AS}-\gamma_\mathrm{DMRG}^\mathrm{AS}|| \; ,
\end{align}
\end{subequations}
and
\begin{subequations}
\begin{align}
    d^\mathrm{b}_\Phi&=||\Phi_\mathrm{CCSD}^\mathrm{AS}-\Phi_\mathrm{DMRG}^\mathrm{AS}|| \; ,\\
        d^\mathrm{b}_\gamma&=||\gamma_\mathrm{CCSD}^\mathrm{AS}-\gamma_\mathrm{DMRG}^\mathrm{AS}|| \; .
\end{align}
\end{subequations}

Note again that in \eqref{eq:dphi_a} the full space CCSD wavefunction has been projected to the active space and normalized to ensure comparability.
At the full space limit, both of them is equal to $d_\Phi$ or $d_\gamma$ value we interested in:
\begin{subequations}
\begin{align}
    d_\Phi&=||\Phi_\mathrm{CCSD}^\mathrm{FS}-\Phi_\mathrm{DMRG}^\mathrm{FS}|| \\
    d_\gamma&=||\gamma_\mathrm{CCSD}^\mathrm{FS}-\gamma_\mathrm{DMRG}^\mathrm{FS}||
\end{align}
\end{subequations}

We can not choose from $d^\mathrm{a}$ and $d^\mathrm{b}$ which will be the better metric, therefore, both will be used. We note, however, that the triangle inequality holds for the 3 distances and thus the difference of $d^\mathrm{a}$ and $d^\mathrm{b}$ is bounded from above by $d^\mathrm{AS}$,

\begin{subequations}
\begin{align}
   |d^\mathrm{a}_\Phi-d^\mathrm{b}_\Phi|<d^\mathrm{AS}_\Phi \; , \\
   |d^\mathrm{a}_\gamma-d^\mathrm{b}_\gamma|<d^\mathrm{AS}_\gamma \; .
\end{align}
\end{subequations}

\begin{figure}
    \centering
    \includegraphics[width=\linewidth]{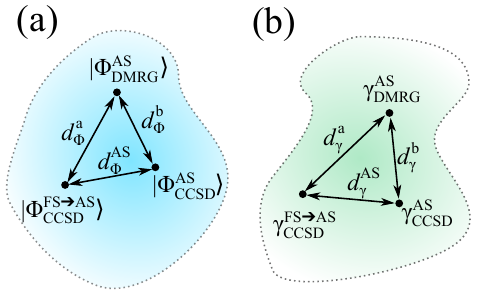}
    \caption{(a) Wavefunctions truncated to single and double excitations on the active space and the distances defined between them. (b) 1RDM-s on the active space and distances defined between them.}
    \label{fig:d_def}
\end{figure}

The relation of various wavefunctions, RDMs and $d$-s introduces in this section are summarized in Fig. \ref{fig:d_def}.
Here we remark, as will be discussed in the following sections, that in practice the average value of $d^a$ and $d^b$ will be used denoted by $\Tilde{d}$.

\subsection{Possible workflows}

\begin{figure}[t!]
\begin{center}
\includegraphics[width=\linewidth]{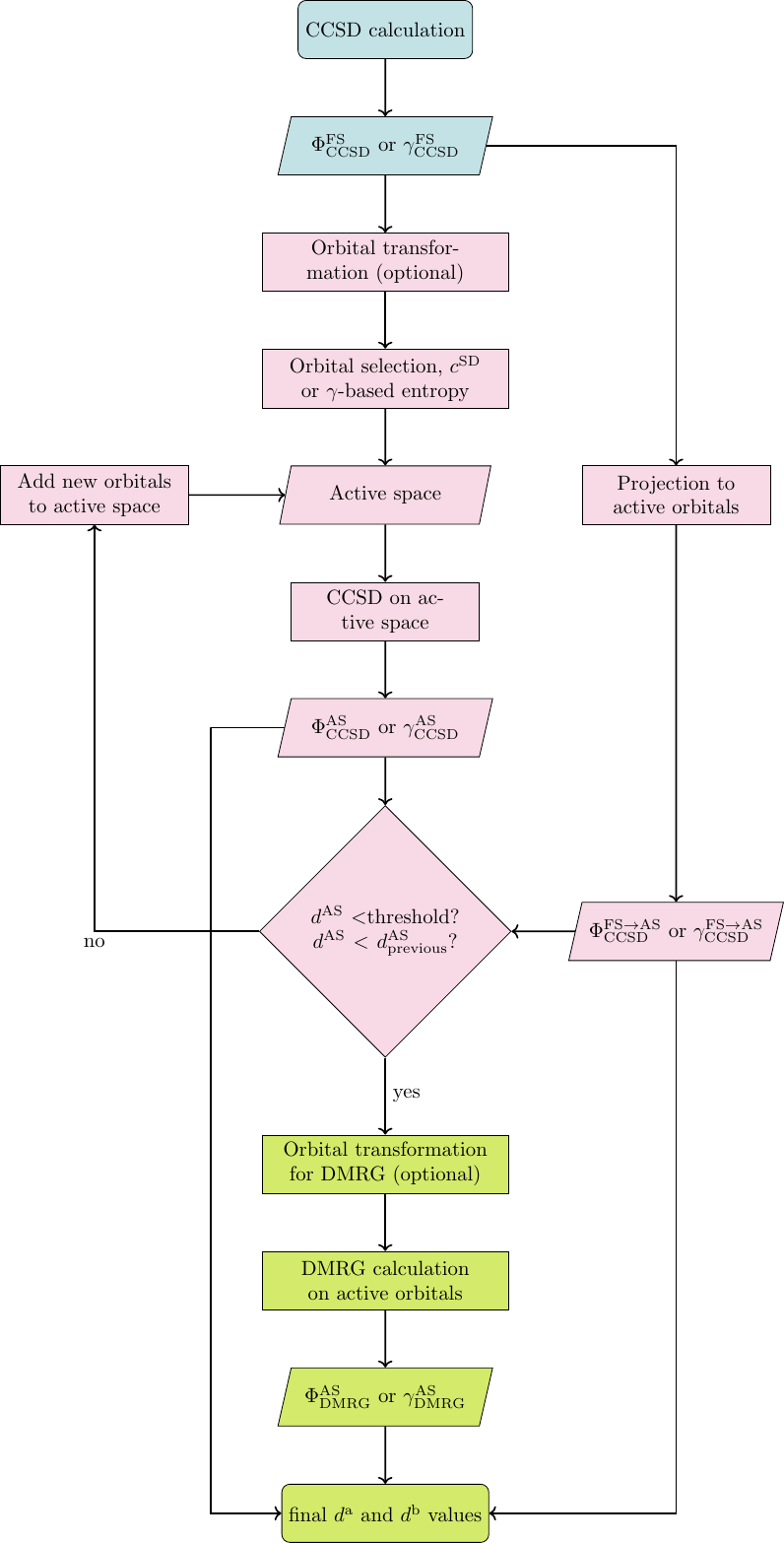}
  \caption{
  Flowchart of the determination of the $d$ diagnostic for a standard CCSD calculation. The different background coloring refers to the different steps, i.e., light blue: original CCSD calculation; light red: active space selection; light green: DMRG calculation and final $d$ determination.
 \label{fig:flowchart}}
 \end{center}
\end{figure}

A general workflow is presented in Fig. \ref{fig:flowchart}:
\begin{enumerate}
\item Perform CCSD calculation and obtain the basis of comparison ($\Phi^\mathrm{FS}_\mathrm{CCSD}$ or $\gamma^\mathrm{FS}_\mathrm{CCSD}$).
\item Perform orbital transformation for better active space selection (optional)
\item Determine orbital entropies based on $c$ or 1- and 2-RDM
\item Select the most influential orbitals
\item Perform CCSD on selected orbital space and obtain $\Phi^\mathrm{AS}_\mathrm{CCSD}$ or $\gamma^\mathrm{AS}_\mathrm{CCSD}$
\item Check if $d^\mathrm{AS}$ is below threshold and smaller than the previous one ($d^\mathrm{AS}_\mathrm{previous}$), if yes proceed, if not repeat from step 4 by including more orbitals until true
\item Perform orbital transformation for DMRG to aid better convergence (optional)
\item Perform DMRG and obtain $\Phi^\mathrm{AS}_\mathrm{DMRG}$ or $\gamma^\mathrm{AS}_\mathrm{DMRG}$
\item Detemine $d^\mathrm{a}$ and $d^\mathrm{b}$
\end{enumerate}

Besides the DMRG accuracy and the $d^\mathrm{AS}$ threshold we have to decide the basis of comparison also if we apply any orbital transformation. The threshold in $d^\mathrm{AS}$ controls the accuracy of the final result while the size of the active space strongly influences the cost of the DMRG calculation. Using some kind of orbital transformation before the orbital selection can compress correlations and can result in smaller active spaces. However, we have to ensure then the invariance of CCSD solution under orbital transformations. The transformation before DMRG, like localization, can lead to better convergence.
Based on the options, we define 3 workflows to apply in this work:

\begin{itemize}
    \item Workflow A: Use $\Phi$ as a basis of comparison, no orbital transformation before selection, split-localization before DMRG. The corresponding result is denoted by $d_\Phi$.
    \item  Workflow B: Use $\gamma$ as a basis of comparison, no orbital transformation before selection, split-localization before DMRG. The corresponding result is denoted by $d_{\gamma}$.
     \item  Workflow C: Use $\gamma$ as a basis of comparison, use "split-natural" orbitals for selection, no additional transformation before DMRG. The corresponding result is denoted by $d_{\gamma,\mathrm{NO}}$.
\end{itemize}

Workflow A is intended to apply when producing RDM-s would be too expensive. Workflow B is the same as A, but with $\gamma$ instead of $\Phi$, because it contains higher excitations from the wavefunction. Workflow C utilizes the fact that if the one-body RDM is at hand, then natural orbitals can be obtained. To keep the invariance of CCSD, the occupied and virtual orbitals should be transformed separately also know as quasi-natural orbitals\cite{Khedkar2019}. We expect that the usage of natural orbitals compresses the correlation and, consequently, less orbitals are needed than in workflow B. In this case, we do not use localized orbitals, because DMRG works well with natural orbitals too~\cite{Fertitta-2014,Ding-2023}.

Importantly, even though the black-box computation of $d$ requires two parameters (the $d^\mathrm{AS}$ threshold and DMRG truncation error) as input, we emphasize that the choice of these parameters is rather flexible. That is, if doubts arise about the reliability of $d$, for example surprisingly high $d$ value (over 0.5) or significantly higher energy with DMRG than CCSD for the same active space, more than 0.01 $E_h$ a more accurate $d$ can be computed at a larger active space and/or a lower truncation error.

Finally, we note that in the workflow both CCSD and DMRG can be substituted to any other method from which CI coefficients or RDM can be extracted to estimate the distance between the wavefunctions of two methods.

\section{Computational details}
\label{sec:Computational_details}

In this section, we present computational details.
Geometries for the W4-17 dataset have been taken from the Supplementary Information of Ref. \citenum{Karton-2017}, for the 3d-MLBE20 dataset from Ref. \citenum{Xu-2015}. Furthermore, nine 3d and nine 4d transition metal species were studied from the work of Bradley et al. \cite{Bradley-2021} using geometries from Refs. \citenum{Merriles-2021-borides,Sun-2009,Anderson-1987,Gutsev-2003,Denis-2005,Suo-2007,Denis-2006}. These species will simply be referred to as 3d-TM9 and 4d-TM9.

As an additional challenge, we collected additional 3d and 4d species from the works of Jiang et al.\cite{Jiang-2012} and Wang et al.\cite{Wang-2015}, which are claimed to be MR. Additionally, based on the work of Sü{\ss} and coworkers\cite{Suss-2020} species also added from the database of Aoto et al.\cite{Aoto-2017}, which have at least 2.5 kcal/mol correction to their bond dissociation energy with icMR-CCSD(T) compared to CCSD(T), along with the additional twelve systems studied by Sü{\ss} et al. Geometries were optimized at B3LYP/cc-pVTZ level of theory, except for species from the database of Aoto et al.\cite{Aoto-2017}. This collection of systems will be labeled as TM-MR. The full list of studied systems can be found in detail in Table S1.

CCSD\cite{Purvis-1982} and CCSD(T)\cite{Raghavachari-1989} calculations were performed using cc-pVDZ~\cite{Dunning-1989,Woon-1993,Balabanov-2005} basis set and frozen core approximation without density fitting with MRCC~\cite{Mester-2025}. A slight modification was made to obtain the CI coefficients from MRCC. Integrals for the subsequent DMRG calculation were also produced with MRCC. For open-shell species ROHF formalism was used.

The occupied and virtual orbitals were localized by Multiwfn~\cite{Lu-2012} applying the Pipek-Mezey localization scheme~\cite{Pipek-1989}.

All DMRG calculations were carried out with the Budapest-DMRG~\cite{budapest_qcdmrg} program package via the DBSS formalism~\cite{Legeza-2003}, by setting the maximum truncation error,
$\epsilon$, to $10^{-5}$.
The minimum value for the bond dimension was set to $D_{\rm min}=256$ and the maximum value was limited by $D_{\rm max}=4096$.
For large active spaces a DMRG orbital ordering was also utilized via a combination of the Fiedler-vector and genetic algorithm based protocols~\cite{Barcza-2011} performing
a series of low-cost DMRG calculations with fixed $D$ = 32, 64, 128 bond dimension values.

The various multireference diagnostics were already discussed in the Introduction; therefore, instead of the detailed description of each metric used in this work, we summarized them in Table \ref{tab:MR_diagnostics}.

Finite-temperature DFT calculations were performed with the default settings of the ORCA (version 5.0.3.~\cite{Neese-2020}) program. All other DFT calculations (BLYP\cite{Becke-1988,Lee-1988}, B1LYP\cite{Adamo-1997}, PBE\cite{Perdew-1996}, PBE0\cite{Adamo-1999}) were performed with MRCC.

The whole workflow is controlled by an in-house developed Python code. Additional MR diagnostics were also evaluated by Python scripts processing DMRG and CCSD CI coefficients, DMRG and CCSD RDMs to obtain occupation numbers along with FT-DFT fractional occupation number, CCSD amplitudes and necessary energies for energy-based diagnostics.

To better understand the practical relationship between the metrics, we determined the pairwise correlation coefficients between different datasets. The commonly used Pearson correlation (r), which lacks robustness and can only reveal linear correlations, is not the most suitable statistic to compare the metrics studied. Correspondingly, in our study Spearman ($\rho$) and Kendall rank correlations were used: ($\tau$)
\begin{align}
    \rho=\frac{\mathrm{cov(R[x][R[y])}}{\rho_{\mathrm{R[x]}}\rho_\mathrm{R[y]}}
\end{align}
\begin{align}
    \tau=\frac{2}{n(n-1)}\sum_{i<j}\mathrm{sgn}(x_i-x_j)\mathrm{sgn}(y_i-y_j)
\end{align}
where $R[x]$ and $R[y]$ are the ranks of $x$ and $y$ variables, and $n$ is the number data points.
They are more reasonable statistics\cite{hollander2013nonparametric}, which uses the ranking of the values and are more robust to outliers. Kendall correlation separates the various metrics even better. In short, we will refer to these correlation coefficients as $r$, $\rho$, and $\tau$, respectively.

\section{Results and discussion}
\label{sec:Results_and_discussion}

In order to keep the focus on the practical application of $d$, in this section only the main summary of the validation protocol is discussed, while details can be found in the SI.

The first approximation is the truncation of the $\Psi$ wavefunction up to double excitations ($\Phi$). The comparison of $d_\Psi$ and $d_\Phi$ shows that $d_\Phi$ holds its descriptive power, although slightly lower than $d_\Psi$.

The next approximation is the usage of DMRG instead of FCI as an FCI solver. By varying the truncation error we found that at least a truncation error of $10^{-5}$ is needed for reasonable results.

The third approximation is the usage of a subspace instead of the whole orbital set. In that sense, we checked the convergence of $d^\mathrm{AS}$, $d^a$ and $d^b$ with respect to the active space size. In general, larger active space leads to lower $d^\mathrm{AS}$, i.e. better quality. $d^a$ and $d^b$ converges smoothly to the full space $d$ in most cases, but, it varies species by species which of them is more useful. However, we noticed, that their average, labeled as $\Tilde{d}$, has better convergence than $d^a$ and $d^b$ in every case, therefore $\Tilde{d}$ will be used to estimate $d$.
Another observation is that NO-transformation can lead to smaller $d$-s by introducing a bias to the CCSD solution. These findings are true for both $\Phi$- and $\gamma$-based protocols.

Based on these results, in the following, we will use two settings, a looser threshold of 0.1 and a tighter 0.05 for both $d^\mathrm{AS}_\Phi$ and $d^\mathrm{AS}_\gamma$. The corresponding results will be noted as $\Tilde{d}_\Phi(0.1)$, $\Tilde{d}_\Phi(0.05)$, $\Tilde{d}_\gamma(0.1)$, $\Tilde{d}_\gamma(0.05)$, where the number in the paranthesis is the used $d^\mathrm{AS}$ threshold. An additional restriction is that the active space selection error should be lower than with the previous active space candidate, to avoid a selection of active space before a local maximum.

\subsection{Comparison of $\Tilde{d}$ and  MR diagnostics on W4-17 dataset}
\label{sec:Results_and_discussion_W4-17}

Before discussing the various $\Tilde{d}$ values, take a look at the performance active space selection schemes. Using the looser 0.1 threshold obviously leads to smaller active spaces. As $\Tilde{d}_\gamma$ is usually higher than $\Tilde{d}_\Phi$, to reach the same threshold on the same set of canonical orbitals, RDM-based orbital selection leads to larger active spaces, 43\% and 37\% of the whole orbital space on average with a threshold of 0.05 for $d^\mathrm{AS}_\gamma$ and $d^\mathrm{AS}_\Phi$, respectively. However, as we seen earlier the usage of NO-transformation leads to less selected orbitals, 34\% of the whole space on average if the threshold is 0.05. Note that these statistics are only valid for this dataset, with larger basis sets we expect smaller ratios.

\begin{table*}[]
\centering
\centerline{\includegraphics[width=1.25\textwidth]{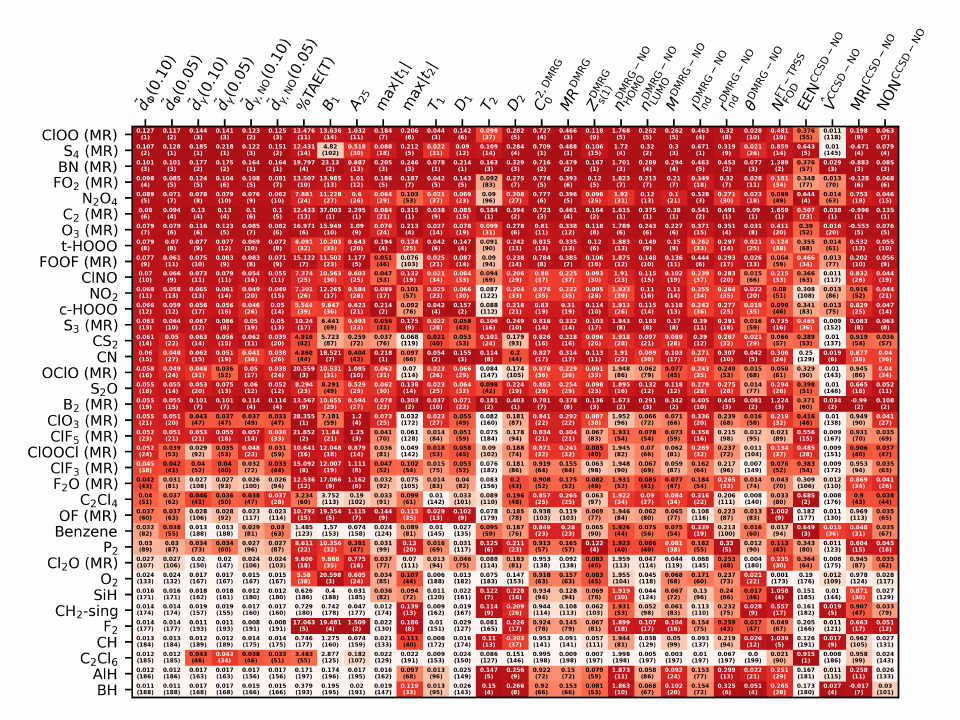}}
    \caption{$\Tilde{d}$ and other commonly used MR diagnostics for a selected subset of W4-17 dataset. It contains the W4-17-MR subset, marked with "MR", and additional species which has a $\Tilde{d}$ value in the top 15 or other metric which is in the highest 5 among the species of W4-17. The number in the cells shows the value of the corresponding metric, while in parentheses the rank of this value among the species is presented. For better visibility, the rankings are also represented by the cell color, where the deeper red color means higher ranking.}
    \label{tab:w4-17mr}
\end{table*}

Instead of discussing the determined $\Tilde{d}$ values for the W4-17 dataset alone, we will analyze them in relation to other existing MR diagnostics, which are listed in Table \ref{tab:MR_diagnostics}.

We think that both MR diagnostics and our $\Tilde{d}$ try to give guidance for the same problem; can we use a single reference method for a given system to obtain meaningful results?
The difference is that MR diagnostics measure the MR character of the studied system, which is usually a metric of how different some property is from a reference, which usually is the ideal single reference wavefunction.
In contrast, $\Tilde{d}$ takes a more practical approach and estimates the distance from the (approximate) FCI wavefunction.
In line with their definition and philosophy, we expect a low correlation between MR diagnostics and $\Tilde{d}$, but it is still worth comparing them since they deal with the same problem.

The quantum chemical method based on which we obtained the given diagnostic is presented in the corresponding superscript. For $C_0^2$ and $MR$ we used the DMRG result instead of performing CASSCF. For metrics based on occupation numbers, CCSD- and DMRG-based RDMs were utilized to derive natural occupation numbers. Alternatively, fractional occupational numbers were also used from finite-temperature DFT.  max$|t_1|$, max$|t_2|$, $T_1$, $D_1$, $T_2$, $D_2$, are determined from CCSD/cc-pVDZ calculations, \%TAE(T) from CCSD(T)/cc-pVDZ, $B_1$ from BLYP/cc-pVDZ and B1LYP/cc-pVDZ, and $A_{25}$ from PBE and PBE0 with cc-pVDZ, respectively.

The various MR diagnostic values for the MR-subset are presented in Table \ref{tab:w4-17mr}. In cases where the definition allows multiple quantum chemical methods (e.g. metrics based on occupation numbers), we chose to show here the ones which are the closest to the definition in the original papers, while other versions can be found in the SI. For demonstrative purposes, we show herein all species where either $\Tilde{d}$ values are in the top 15 or one of the other diagnostics listed have a top 5 value among the species of the W4-17 dataset. The threshold when a species is labeled as multireference is not always defined and is sometimes debatable. Although we label the members of the "MR" subset, as given in the W4-17 paper~\cite{Karton-2017}, in Table \ref{tab:w4-17mr} next to the molecular formula, we do not consider the W4-17-MR molecules as a specific group in our discussion. Instead, we rank the molecules of the W4-17 dataset based on each investigated diagnostic value, and use the ranking numbers as guidelines. This ranking is shown in parenthesis in Table \ref{tab:w4-17mr} and is also visualized by the coloring.

\begin{table*}
    \centering
    \centerline{\includegraphics[width=1.2\textwidth]{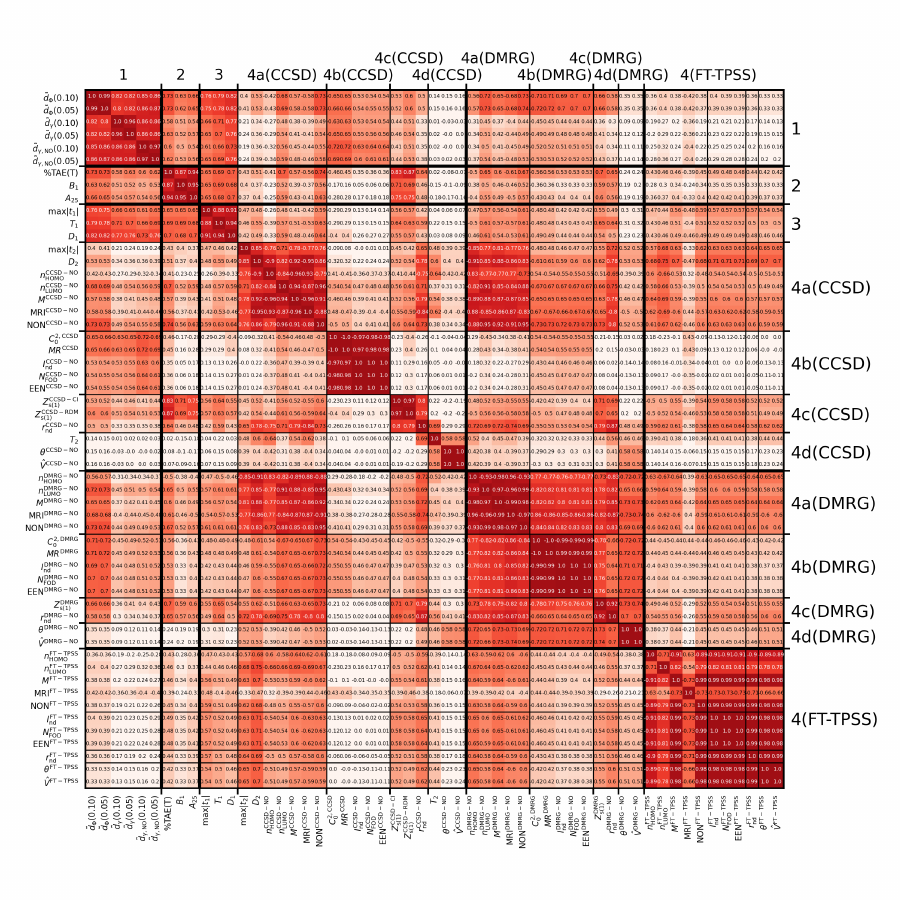}}
      \caption{Pairwise Spearman rank correlation matrix of $\Tilde{d}$ and other MR diagnostics for the W4-17 dataset. The more intense red hue represents higher correlation. The left and lower labels show the metrics, while the right and upper labels show the corresponding groups.}
    \label{fig:heatmaps}
\end{table*}

\begin{table*}
    \begin{tabular}{c|c|c|cccc}
      1  &    2      &     3       & 4a & 4b & 4c & 4d \\
    \hline

    \hline
    $\Tilde{d}_\Phi$ & \%TAE(T) & max|$t_1$| & max|$t_2$|  & $C_0^2$  & $Z_\mathrm{s(1)}$ & $T_2$ \\
    $\Tilde{d}_\gamma$& $B_1$    & $T_1$      & $D_2$            & $MR$              & $r_\mathrm{nd}$ &  $\theta$ \\
    $\Tilde{d}_{\gamma,\mathrm{NO}}$    & $A_{25}$ & $D_1$      & $n_\mathrm{HOMO}$               & $I_\mathrm{nd}$ & & $\hat{V}$ \\
        &          &            & $n_\mathrm{LUMO}$   & $N_\mathrm{FOD}$             & & \\
        &          &            &  $M$              &   EEN             & & \\
        &          &            &  NON    &  &   &    \\
        &          &            &  MRI    & &    &    \\
    \end{tabular}
    \caption{Groupings of MR diagnostics based on Pearson, Spearman, and Kendall $\tau$ correlations of each metrics result on the species of W4-17 dataset.}
\end{table*}

The different $\Tilde{d}$ formulations consistently identify species with high $\Tilde{d}$. The largest discrepancy is found for \ce{B2} where $\Tilde{d}_\Phi$(0.05) gives 0.055, which can be considered a relatively high value among $\Tilde{d}_\Phi$-s, while $\Tilde{d}_\gamma$(0.05) and $\Tilde{d}_{\gamma,\mathrm{NO}}$(0.05) produce 0.101 and 0.114, respectively. Other notable differences are benzene with relatively small $\Tilde{d}_\gamma$ (0.013) and \ce{C2Cl6} with relatively high $\Tilde{d}_\gamma$ (0.043) if compared to $\Tilde{d}_\Phi$ values (0.038 and 0.012) and rankings.

In general, the looser and tighter threshold for active space produces similar results, but the different formulations differ considerably. $\Tilde{d}_\Phi$-s are smaller than $\Tilde{d}_\gamma$ due to the fact that $c$ coefficients are used up to double excitations. Another thing that was noticed earlier was that NO-transformation also leads to smaller $\Tilde{d}$ values, even though we expect similar results. More detailed statistical analysis will be discussed later.

Considering the rest of the metrics, there are numerous discrepancies for the studied species.
From \ce{ClOO} to \ce{B2} (upper half of the table) most diagnostics and $\Tilde{d}$ have high values, but max$|t_2|$, $T_2$, EEN and $\hat{V}$ disagrees with them notably. Regarding the W4-17 classification, which is based on the \%TAE(T) diagnostics, \ce{N2O4}, t-HOOO, c-HOOO, \ce{NO2}, \ce{S2O}, \ce{CS2} and \ce{CN} are species with notable $\Tilde{d}$ (higher than 0.05) but not part of the MR subset. However, \ce{OClO}, \ce{ClOOCl}, \ce{ClF3}, \ce{F2O}, \ce{OF}, and \ce{Cl2O} have all $\Tilde{d}$ (0.05)-s less than 0.05, although they are part of the W4-17-MR subset. Hydrides are interesting cases, because $\Tilde{d}$ and energy-based metrics show low values, but $T_2$, $D_2$, $n_\mathrm{HOMO}$, $M$, $I_\mathrm{nd}$, $r_\mathrm{nd}$, $\theta$, FOD, $\hat{V}$, and MRI shows significant MR character. For energy-based diagnostics \ce{Cl2O}, \ce{O2}, and \ce{F2} seem to be problematic cases.

To better understand the practical relationship between the metrics, we determined the pairwise correlation coefficients ($r$, $\rho$, $\tau$).
Table~\ref{fig:heatmaps}. shows the Spearman rank correlation matrix, while the other two ($r$, $\tau$) can be found in the SI.

First, we look at the correlation between the $\Tilde{d}$ metrics. The correlation between the same metrics with different active space threshold is close to 1 ($r$=0.97-0.98, $\rho$=0.97-099, $\tau$=0.87-0.94). However, between the three different types of $\Tilde{d}$ metrics there is a lower but still strong correlation. $\Tilde{d}_\Phi$-$\Tilde{d}_\gamma$ have $r$=0.86-0.91, $\rho$=0.80-0.82, $\tau$=0.63-0.66; $\Tilde{d}_\Phi$-$\Tilde{d}_{\gamma,\mathrm{NO}}$ have $r$=0.86-0.90, $\rho$=0.85-0.87, $\tau$=0.68-0.70; and $\Tilde{d}_{\gamma}$-$\Tilde{d}_{\gamma,\mathrm{NO}}$ have $r$=0.92-0.96, $\rho$=0.86, $\tau$=0.69-70.

Based on these statistics, the metrics can be categorized into a few groups.
The different $\Tilde{d}$ metrics are in their own category (group 1) and do not correlate with any other metrics, which is understandable because it is totally different from the others and try to answer a different question.
Energy-based descriptors (\%TAE(T), $B_1$, $A_{25}$) are making their own group (group 2). This is not surprising because all of them use total atomization energies.
The metrics derived from single amplitudes (max|$t_1$|, $T_1$, $D_1$) are also separated from the other diagnostics (group 3), which mostly measure orbital relaxation effects. However, for other multireference diagnostics the main distinguishing property is not the formulation but the source of the data, i.e. occupation number and entropy-based diagnostics are grouped by that they derived from CCSD, DMRG or FT-DFT. Within these groups, the metrics can be further divided.

Group 4a contains the simple occupation number-based diagnostics ( $n_\mathrm{HOMO}$, $n_\mathrm{LUMO}$, $M$, NON ), and surprisingly MRI, which is a complex transformation of them. Also, NON which is basically the spinorbital version of $n_\mathrm{LUMO}$.
In group 4b we can see $C_0^2$ and MR is closely related, because $C_0^2$ is the largest contributor of MR. EEN and $N_\mathrm{FOD}$ are also defined in a way that they are identical, only $N_\mathrm{FOD}$ is half of EEN.  $I_\mathrm{nd}$ have also a similar form to $MR$ ($\sum_{\sigma,i} n_i^\sigma(1-n_i^\sigma)=\sum_{\sigma,i} n_i^\sigma - n_i^{\sigma 2} \sim \sum_i |c_i|^2 - |c_i|^4 $).
$\theta$ and $\hat{V}$ in group 4d also have a similar philosophy by taking the difference of occupation numbers from the ideal and scaling with the number of electrons.

Notice that metrics from double amplitudes can also be sorted in this grouping with other diagnostics derived from CCSD natural orbital occupations.
Looking at the off-diagonal elements, we can see that the subgroups of group 4 are fairly correlated for diagnostics derived from DMRG and FT-DFT (except for MRI$^\mathrm{FT-TPSS}$), while in the case of CCSD the subgroups are much more separated, especially group 4b. For DMRG, the reason might be that approaching FCI the differences between the various MR definitions are getting smaller. In case of FT-TPSS data we can observe that the default settings produce occupation numbers that do not differ from the occupation of the reference state, leading to small separation for SR species and large deviation from other metrics.

Group 1 has some noticable correlation only with group 3 ($r$=0.53-0.77, $\rho$=0.61-0.82 $\tau$=0.44-0.63), and $\Tilde{d}_\Phi$ have with DMRG-based group 4a and 4b ($r$=0.67-0.84, $\rho$=0.56-0.74 $\tau$=0.40-0.54).
These are not that high, therefore $\Tilde{d}$ cannot be estimated using other existing MR diagnostics. Another connection can be observed between the members of the CCSD and the DMRG-based subgroup 4a ($r$=0.68-0.95, $\rho$=0.73-0.95, $\tau$=0.55-0.81).

Earlier comparisons found similar conclusions as we did. Fogueri et al.\cite{Fogueri-2013} showed that various $A_\lambda$ and \%TAE formulations correlate well with each other, and $C_0^2$ and $M$ form another group.
Duan and coworkers\cite{Duan-2020} studied 15 MR diagnostics and also concluded that the source of the data for the diagnostics is an important factor. They could also distinguish energy-based descriptors from other metrics, that $T_1$ and $D_1$ behave similarly, and $D_2$ correlates moderately well with other occupation-based diagnostics.
Most recently, Martin et al.\cite{Martin-2022} studied the correlation between various MR diagnostics and derived TAEx. They determined four clusters: (1) $T_1$, $D_1$, max $t_1$; (2a) $r_\mathrm{nd}$, $\overline{I_\mathrm{nd}}$ (divided by the number of electrons, not 2, leading to the same expression as $\hat{V}$); (2b) $M$, $D_2$, max|$t_2$|; (3) energy-based diagnostics. These groupings are in line with our findings.
Xu et al.\cite{Xu-2023} showed a connection between $\overline{I_\mathrm{nd}}$ and $C_0$, and also between $D_2$ and maximum of $n_i^\sigma(1-n_i^\sigma)$ values, labeled as $I_\mathrm{nd}^\mathrm{max}$. We already grouped the unscaled $I_\mathrm{nd}$ and $C_0^2$ together; however, the size effects are less prominent in the used dataset.
In a more recent work\cite{Xu-2025} they compared several MR diagnostics, and they determined five groups of correlation measures. From these, the dynamic natural orbital occupation (NOO)-based correlation measure ($\overline{I_\mathrm{D}}$) is not a good MR diagnostic, while average NOO-based nondynamic correlation measures ($\overline{I_\mathrm{nd}}$ and $\%I_\mathrm{nd}=\overline{I_\mathrm{nd}}/(\overline{I_\mathrm{nd}}+\overline{I_\mathrm{d}})\times100$) could correspond to our 4b group.
Maximal nondynamic correlation measures that include $I^\mathrm{max}_\mathrm{nd}$, $n_H$, $n_L$ (maximum and minimum occupancy of occupied and virtual natural spinorbitals), -log(I) (connected to MRI, defined in Table \ref{tab:MR_diagnostics}), $M$, and $D_2$, is similar to our group 4a. $D_1$ and $T_1$ is in $t_1$-based correlation measures, as in our group 3, and energy-based correlation measures corresponds to our group 2.

Recently, Stanton and his coworkers presented Density Asymmetry Diagnostic (DAD)\cite{Weflen-2025}, which measures the antysymmetry of the 1-RDM of a CC calculaton. In our used code packages (MRCC) such quantity is not accessible and only the real symmetrized part of the 1-RDM is returned. Therefore, we did not determine DAD for the studied species, but we compared our numbers on the W4-11 subset that they used. In this subset DAD belongs to group 3.

Based on these findings, we can conclude that none of the existing MR diagnostics provides information similar to any formulation of $\Tilde{d}$. However, this is not surprising if we consider that our new metric measures a different property, estimating the distance from the exact solution, while the rest of the MR diagnostics measure the deviation from the ideal single reference solution. Following this line of thought, we are not classifying systems as MR or SR based on $\Tilde{d}$, rather evaluating the suitability of the method used, here CCSD. $\Tilde{d}$ increases with the size of the system, and based on the comparison of alkanes and halogenated alkanes (Table S5.) found that this scaling is proportional to the number of non-H or heavy atoms ($N_\mathrm{heavy}$). Using this scaling, if $\Tilde{d}_\gamma$ is below $0.03536\times\sqrt{N_\mathrm{heavy}}$ we consider that the used method is highly reliable, while if $\Tilde{d}$ is over $0.07071\times\sqrt{N_\mathrm{heavy}}$ then the used method is not reliable, based on the numerical results on W4-17 dataset.
Between the two thresholds the used method is moderately reliable. The respective thresholds for $\Tilde{d}_\Phi$ are $0.03\times\sqrt{N_\mathrm{heavy}}$ and $0.06\times\sqrt{N_\mathrm{heavy}}$. We note that these thresholds are arbitrary, as most of them do not have strict criteria, because we do not have another reference on which we can decide the thresholds.
Based on these limits according to $\Tilde{d}_\gamma(0.05)$ the calculation of BN, \ce{S4}, \ce{C2}, \ce{ClOO}, \ce{B2}, and \ce{O3} are problematic, and \ce{FO2}, \ce{S3}, \ce{ClNO}, t-HOOO, \ce{S2O}, FOOF, \ce{CN}, and \ce{NO2} are moderately difficult for CCSD.

We also note here the differences between the different $\Tilde{d}$-s. Generally $\Tilde{d}_\Phi$-s are smaller than $\Tilde{d}_\gamma$, but both have the predictive power to estimate the difference from the FCI. $\Tilde{d}_{\gamma,\mathrm{NO}}$ should lead to a similar value as $\Tilde{d}_\gamma$, but with fewer selected orbitals, however, in some cases NO-transformation introduces a bias to the original CCSD solution, leading to a more similar DMRG solution and smaller $\Tilde{d}_{\gamma,\mathrm{NO}}$.

\begin{table*}
\begin{center}

\caption{\label{tab:W4-17_CCSDT} $\Tilde{d}$ values for species of W4-17 with $\Tilde{d}_\gamma(0.05)$ over 0.075 for CCSD/cc-pVDZ calculations.}
\centerline{
\scalebox{0.9}{
\begin{threeparttable}

 \begin{tabular}{c|ccc|ccc|ccc}
  & \multicolumn{3}{c}{CCSD/cc-pVDZ} & \multicolumn{3}{c}{CCSD/cc-pVTZ} & \multicolumn{3}{c}{CCSDT/cc-pVDZ} \\
Species  & $\Tilde{d}_\Phi(0.05)$ & $\Tilde{d}_\gamma$(0.05) & $\Tilde{d}_{\gamma,\mathrm{NO}}$(0.05) & $\Tilde{d}_\Phi(0.05)$ & $\Tilde{d}_\gamma$(0.05) & $\Tilde{d}_{\gamma,\mathrm{NO}}$(0.05) & $\Tilde{d}_\Phi(0.05)$ & $\Tilde{d}_\gamma$(0.05) & $\Tilde{d}_{\gamma,\mathrm{NO}}$(0.05)\\
\hline
\ce{B2}   & $0.055$ & $0.101$ & $0.114$ &
$0.053$ & $0.121$ & $0.125$ &
$0.013$ & $0.023$ & $0.033$ \\
\ce{BN}   & $0.101$ & $0.175$ & $0.164$ &
$0.106$ & $0.195$ & $0.170$ &
$0.025$ & $0.028$ & $0.029$ \\
\ce{C2}   & $0.094$ & $0.130$ & $0.100$ &
$0.096$ & $0.165$ & $0.149$ &
$0.021$ & $0.026$ & $0.026$ \\
\ce{ClNO} & $0.066$ & $0.079$ & $0.055$ &
$0.065$ & $0.072$ & $0.057$ &
$0.028$ & $0.027$ & $0.031$ \\
\ce{ClOO} & $0.117$ & $0.141$ & $0.125$ &
$0.111$ & $0.140$ & $0.130$ &
$0.027$ & $0.026$ & $0.022$ \\
\ce{FO2}  & $0.085$ & $0.104$ & $0.081$ &
$0.084$ & $0.103$ & $0.091$ &
$0.028$ & $0.028$ & $0.025$ \\
\ce{FOOF} & $0.061$ & $0.083$ & $0.071$ &
$0.067$ & $0.095$ & $0.066$ &
$0.029$ & $0.036$ & $0.029$ \\
\ce{N2O4} & $0.071$ & $0.079$ & $0.062$ &
$0.071$ & $0.081$\tnote{a} & $0.062$ &
$0.034$ & $0.038$ & $0.047$ \\
\ce{O3}   & $0.079$ & $0.123$ & $0.082$ &
$0.082$ & $0.138$ & $0.083$ &
$0.026$ & $0.031$ & $0.028$ \\
\ce{S2O}  & $0.055$ & $0.075$ & $0.052$ &
$0.062$ & $0.064$ & $0.049$ &
$0.026$ & $0.032$ & $0.026$ \\
\ce{S3}   & $0.064$ & $0.086$ & $0.050$ &
$0.075$ & $0.105$ & $0.049$ &
$0.022$ & $0.036$ & $0.027$ \\
\ce{S4}   & $0.128$ & $0.218$ & $0.151$ &
$0.122$\tnote{b} & $0.221$\tnote{a} & $0.158$ &
$0.032$ & $0.059$ & $0.059$ \\
t-HOOO    & $0.070$ & $0.077$ & $0.072$ &
$0.071$ & $0.076$ & $0.070$ &
$0.022$ & $0.028$ & $0.026$ \\

\end{tabular}
\begin{tablenotes}
\item[a] $\Tilde{d}_\gamma(0.07)$
\item[b] $\Tilde{d}_\Phi(0.07)$
 \end{tablenotes}
\end{threeparttable}}}
\end{center}
\end{table*}

We close this section by emphasizing that the $\Tilde{d}$ diagnostic values are clearly attached to the studied calculation.
To highlight this important feature, we determined the same $\Tilde{d}(0.05)$ metrics but with CCSD/cc-pVTZ and CCSDT/cc-pVDZ calculation for systems which have $\Tilde{d}_\gamma(0.05)$ over 0.075 (Table \ref{tab:W4-17_CCSDT}) to see the effect of basis set and method. As expected, really similar values have been obtained with the larger cc-pVTZ basis set, as with the smaller cc-pVDZ. As for the method, it is clear that eventhough CCSD might be inappropriate in some cases, higher-level methods, like CCSDT, have no problem with the species in question. This is demonstrated by the low (<0.05) values of $\Tilde{d}(0.05)$ for all species and formulations, except \ce{S4} which have a $\Tilde{d}_\gamma$(0.05)=0.059 with CCSDT, which is still significantly lower than 0.218 for CCSD. This finding also proves the applicability of high-accuracy thermochemical models like W$n$, HEAT, cccA, etc. for the W4-17 dataset.

\subsection{Application of the $\Tilde{d}$ metric to transition metal species}
\label{sec:Results_and_discussion_metals}

\begin{figure*}[t]
\begin{center}

       \includegraphics[width=\textwidth]{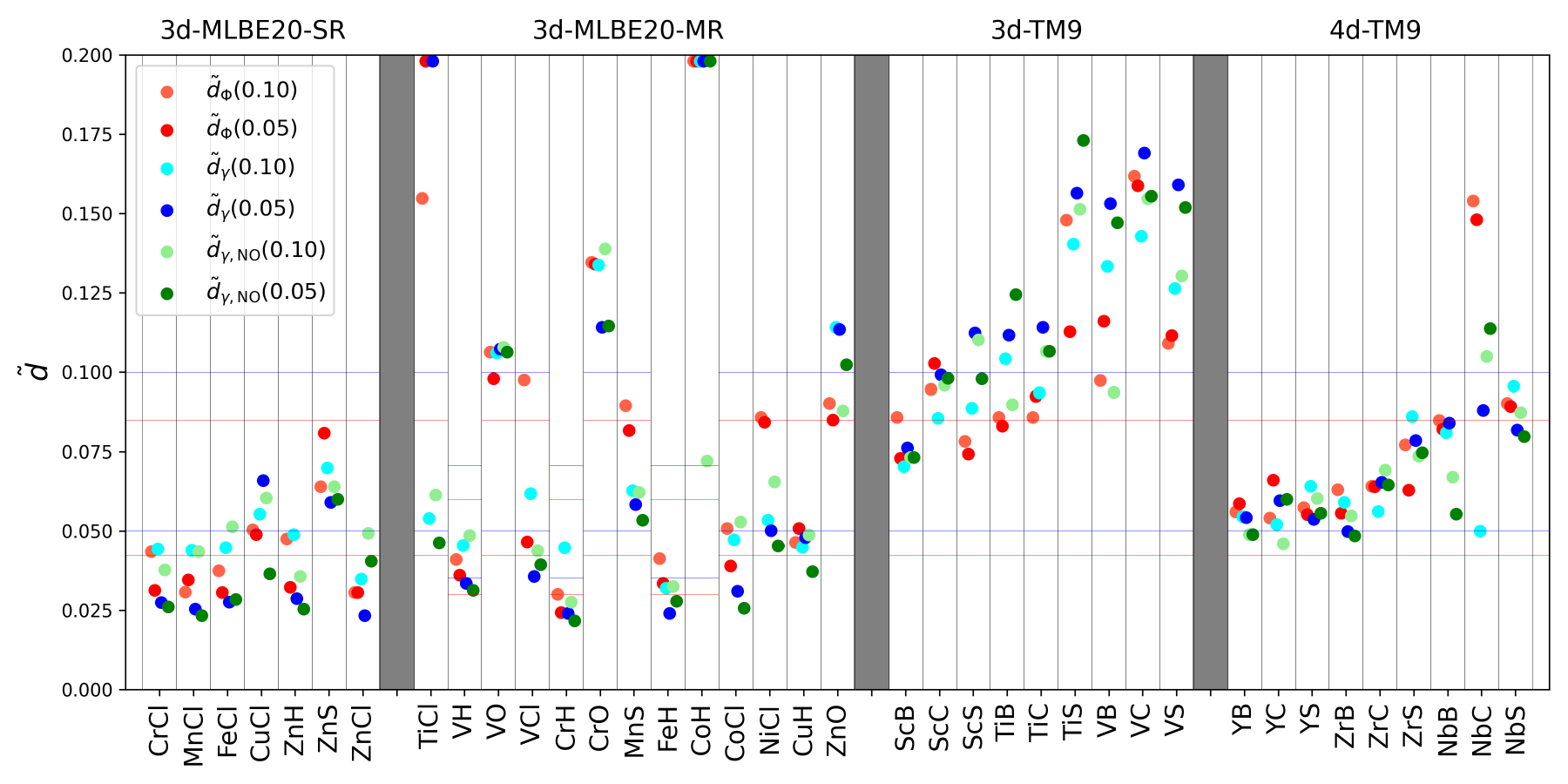}
       \includegraphics[width=\textwidth]{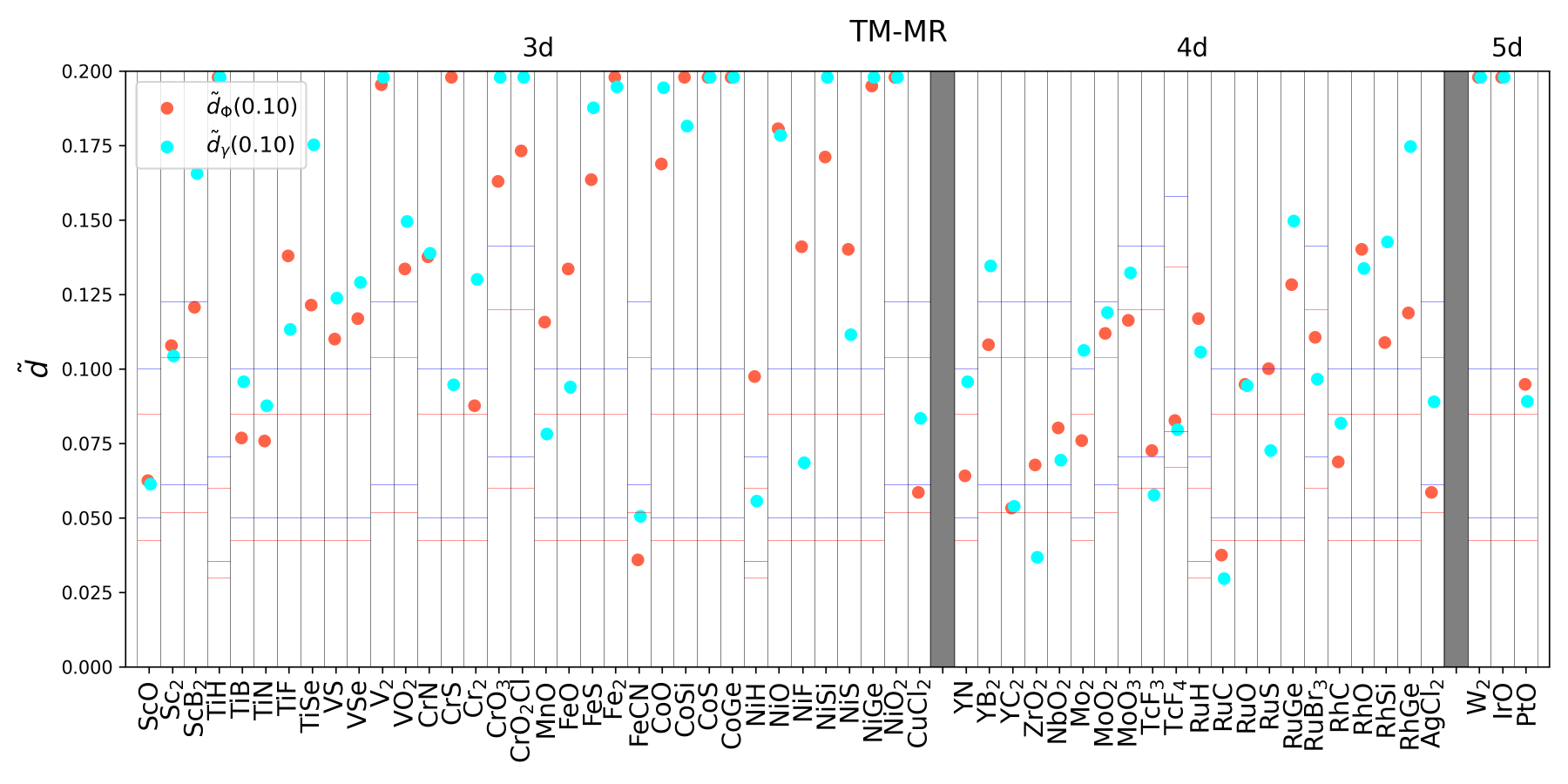}
  \caption{$\Tilde{d}$ values for TM species studied in this work. Horizontal red and blue lines represent the lower and upper thresholds for $\Tilde{d}_\Phi$ and $\Tilde{d}_\gamma$ values, respectively. }
 \label{fig:TM}
 \end{center}
\end{figure*}

The multireference character among the transition metal species is much more common due to the several low-lying states. In these cases, choosing a reliably accurate yet numerically affordable approach can be more problematic. For main group species, the single reference methods usually give correct results, even sub-kJ/mol accuracy is achievable. However, the usage of multireference methods is more complicated and there is no gold standard such as CCSD(T) for single reference systems yet.

In such challenging situations, the usage of $\Tilde{d}$ metrics can be even more helpful to decide if single reference method is applicable to a transition metal species in question. We demonstrate the concept by applying the $\Tilde{d}$ metric for the 3dMLBE dataset\cite{Xu-2015} and for species studied with s-ccCA recently\cite{Bradley-2021} labeled as 3d-TM9 and 4d-TM9 for 3d and 4d transition metal species, respectively.
For comparison, results based on $\Tilde{d}$ analysis are presented in Fig. \ref{fig:TM}.
The 3d-MLBE20 dataset is divided into two subsets where 7 species are considered as single reference and 13 species are considered MR. In case of the 3d-MLBE20-SR subset $\Tilde{d}$-s are fairly low, mostly below 0.05, only CuCl and ZnS show moderately high values, $\Tilde{d}_\gamma(0.05)$ is 0.066 and 0.059, respectively.

From the 3d-MLBE20-MR subset, VH, VCl, CrH, FeH, CoCl, and CuH have all $\Tilde{d}(0.05)$-s below the corresponding threshold showing that CCSD is adequate for studying these systems, although it is in the MR subset. TiCl and CoH have exceptionally high $\Tilde{d}_\gamma$(0.05) (0.427 and 0.305), but in the case of TiCl $\Tilde{d}_{\gamma,\mathrm{NO}}(0.05)$ it completely disagrees (0.045). This is the most spectacular case where the NO-transformation causes such a deviation. VO, CrO, and ZnO have also relatively high $\Tilde{d}$-s, mostly over 0.1.

Looking at 3d-TM9 and 4d-TM9, in Ref.~\citenum{Bradley-2021} the multireference character was classified according to the criteria of $T_1$ > 0.05/0.045, $D_1$ > 0.15/0.12 and \%TAE(T) > 10\%/10\% for 3d and 4d species, respectively. Based on this metric, seven 3d species (ScB, ScC, ScS, TiB, VB, VC, VS) and zero 4d species are considered to have multireference character. However, every $\Tilde{d}$ is greater than 0.05, suggesting that at most only moderately adequate to use CCSD for them. 3d-TM9 species are especially challenging; there are only two species with $\Tilde{d}_\gamma(0.05)$ below 0.1, ScB (0.076) and ScC (0.099). 4d-TM9 species are mostly between 0.05-0.10, although clasified as single reference, NbC even shows a $\Tilde{d}_\Phi(0.05)$ of 0.148.

As an additional challenge we determined $\Tilde{d}_\Phi(0.10)$ and $\Tilde{d}_\gamma(0.10)$ values for species, labeled as TM-MR, selected from Refs \citenum{Jiang-2012,Wang-2015,Suss-2020} which are claimed to be MR systems. Based on the size-dependent thresholds determined in the previous section FeCN, \ce{YC2}, and RuC have low $\Tilde{d}(0.1)$ values suggesting that CCSD is adequate for them. For ScO, TiB, TiN, \ce{CuCl2}, YN, \ce{ZrO2}, \ce{NbO2}, \ce{MoO3}, \ce{TcF3}, \ce{TcF4}, \ce{RuBr3}, RhC, and \ce{AgCl2} both $\Tilde{d}$ variant is between their respective thresholds, meaning that one should be careful using CCSD.
Some systems have differing $\Tilde{d}_\Phi(0.10)$ and $\Tilde{d}_\gamma(0.10)$ values, placing them on different sides of the higher threshold. These species are \ce{CrS}, \ce{Cr2}, \ce{MnO}, \ce{FeO}, NiH, NiF,  \ce{Mo2}, \ce{MoO2}, RuO, RuS, and PtO.
The rest of the species are problematic for CCSD, especially TiH, \ce{V2}, \ce{CrO3}, \ce{CrO2Cl},  \ce{Fe2}, CoS, CoGe, CoSi, NiSi, NiGe, \ce{NiO2}, \ce{W2}, and IrO for which one of the $\Tilde{d}(0.01)$-s is over 0.2.

Note that, in line with the previous findings,\cite{Hait-2019,Cheng-2017,Fang-2017} our results show that even if a system is regarded as multireference by some criteria, it may be handled with single reference CC to provide reliable data.

Regarding the composition of active spaces, we found high variance. For simpler systems the occupied $d$ and a few virtual $d$ orbitals with low principal numbers of a given TM and some occupied and virtual $s$ and $p$ orbitals from the ligand is enough based on convergence, but the proper description of more correlated systems asks for orbitals with even higher principal numbers (4$d$, 5$d$, etc.) both for the TM and non-TM atoms.

Before concluding this section, we want to highlight another application of the framework presented in this paper. With DMRG the different excited states can be easily determined, which is useful especially in case of transition metal species to determine the correct ground state which can be problematic to find. Usually within this framework, such problems are highlighted with extremely high $\Tilde{d}$ values, over 1. In such cases, asking for more eigenvalues from the DMRG calculation and looking at the most dominant determinants can help to identify the correct ground state.
In this work, we found that the ground state of FeH is not $^4\Delta$, but $^6\Delta$, of CoCl not $^3\Phi$, but $^5\Delta$, YC is not $^4\Pi$, but $^4\Sigma$, and \ce{YC2} is not $^2$B$_2$, but $^2$A$_1$. Note that this is only true for cc-pVDZ basis set, it is possible that the order of these closely lying states are changing with increased basis sets. However, such cases require a more careful approach and study the possible occupations.

\section{Conclusion}
\label{sec:Conclusion}

In this work, we present a new strategy to determine the quality of post-HF wavefunctions. Instead of trying to measure the deviance from the ideal single reference picture as many multireference diagnostics do, we measure the difference from an approximate FCI solution. Here, we focus on CCSD to demonstrate the applicability of the workflow. To this end, a subset of orbitals is chosen based on the CCSD solution, and a DMRG calculation is performed on them. Then the wavefunctions up to double excitations, $\Phi$ or 1-body RDM-s, $\gamma$, are compared from DMRG and CCSD to give a single metric labeled as $\Tilde{d}_\Phi$ and $\Tilde{d}_\gamma$.

We also introduced an orbital selection procedure that ensures that the selected subspace correctly represents the entire space. During this procedure, the orbitals are ranked by their relative importance derived from the CCSD solution, and on the selected subspace another CCSD is performed and the $\Phi$-s or $\gamma$-s are compared to the original CCSD solution. The subspace is increased until the two CCSD solution is similar enough. This method enables us to tune the accuracy of $\Tilde{d}$.
The usage of natural orbitals from the CCSD before orbital selection was also tested, assuming it helps to select smaller subspaces.

Three different workflows were tested on the W4-17 dataset and compared with other popular multireference diagnostics. The three flavors of $\Tilde{d}$ correlate well with each other ($\Tilde{d}_\Phi$, $\Tilde{d}_\gamma$, $\Tilde{d}_{\gamma,\mathrm{NO}}$). $\Tilde{d}_\Phi$ is usually smaller than $\Tilde{d}_\gamma$, therefore, with the same active space selection threshold $\gamma$-based selections are somewhat larger. However, NO-transformation produces smaller active spaces, but also introduces a bias to the CCSD solution that leads to smaller $\Tilde{d}_{\gamma,\mathrm{NO}}$ than $\Tilde{d}_\gamma$.

None of the metrics was found to correlate well with $\Tilde{d}$-s. MR diagnostics were also grouped by their performance and found that the source of data is as important as the formulation of the MR diagnostic itself. The following groups were determined, which are in line with previous findings: group 2: \%TAE(T), $B_1$, $A_{25}$ ; group 3: max|$t_1$|, $T_1$, $D_1$; group 4a: $n_\mathrm{HOMO}$, $n_\mathrm{LUMO}$, $M$, MRI, NON, max|$t_2$|, $D_2$; group 4b: $C_0^2$, $MR$, $I_\mathrm{nd}$, $N_\mathrm{FOD}$, EEN; group 4c: $Z_\mathrm{s(1)}$, $r_\mathrm{nd}$; group 4d: $\theta$, $\hat{V}$ $T_2$.

The proposed approach was also tested on transition metal species. VH, VCl, CrH, FeH, CoCl, and CuH have $\Tilde{d}$-s below 0.05 with CCSD, showing that it is an adequate method to use on them, despite being in the MR subset. All 3d-TM9 and 4d-TM9 species have $\Tilde{d}_\gamma$(0.05)-s over 0.05 highighting, that these are at least moderately challenging. Additionally, 3d species are above 0.1, with the exception of ScB, therefore, CCSD is not recommended to study them.
For a selection of MR transition metal species we found that the calculation of FeCN, \ce{YC2}, and RuC are not problematic and ScO, TiB, TiN, \ce{CuCl2}, YN, \ce{ZrO2}, \ce{NbO2}, \ce{MoO3}, \ce{TcF3}, \ce{TcF4}, \ce{RuBr3}, RhC, and \ce{AgCl2} are only moderately difficult for CCSD.

With regard to FeH, CoCl, YC, and \ce{YC2} molecules we predicted ground states different from those of the literature findings; nevertheless, these results are inconclusive because of the potential basis set sensitivity of the problems. However, our results demonstrate that the proposed procedure is also capable of finding the correct ground state, which is not always straightforward in transition-metal chemistry.

Based on the results, we find that in the case where $\Tilde{d}_\gamma$ ($\Tilde{d}_\Phi$) is below $0.03536\times\sqrt{N_\mathrm{heavy}}$ ($0.03\times\sqrt{N_\mathrm{heavy}}$), the method used is adequate, while for $\Tilde{d}_\gamma>0.07071\times\sqrt{N_\mathrm{heavy}}$ ($\Tilde{d}_\Phi>0.06\times\sqrt{N_\mathrm{heavy}}$) the used method is not reliable and other approaches are suggested to study the system in question. In the intermediate region of $\Tilde{d}$, one should be cautious with the derived results. It must be noted that these limits are not strict just recommendations based on the results in this work.

The approach presented here can be applied to test the quality of any wavefunction method where CI coefficients or RDMs can be extracted and a proper active space can be selected. In the matter of high-level reference methods, DMRG can be substituted for any reference-free method that is capable of providing an approximate FCI solution for large enough active spaces. This way the approach has the potential to become a quality assurance tool for wavefunction based methods.

\section{Supporting Information}
Geometries of the studied species are provided in a zip file.
Results for W4-17 and transition metal species, MR diagnostics for W4-17, and Pearson, Spearman, Kendall correlation statistics between the diagnostics are given in a spreadsheet.
Validation of the introduced approximations, additional figures and tables presented in a pdf file.

\section{Acknowledgments}
This work was supported by the Ministry of Innovation and Technology and the National Research, Development and Innovation Office of Hungary (NKFIH) within the National Quantum Technology Program (Grant No. 2022-2.1.1-
NL-2022-00004) and Grant Nos.~K134983, TKP2021-NVA-04, PD-146265, FK-135496.
\"O.L. acknowledges support of the Hans Fischer Senior Fellowship programme funded by the Technical
University of Munich - Institute for Advanced Study, and of the Center
for Scalable and Predictive methods
for Excitation and Correlated phenomena (SPEC), funded as part of the
Computational Chemical Sciences Program FWP 70942 by the U.S. Department of
Energy (DOE), Office of Science, Office of Basic Energy Sciences, Division
of Chemical Sciences, Geosciences, and Biosciences at Pacific Northwest
National Laboratory. Z.B. acknowledges the financial support
of the János Bolyai Research Fellowship of the Hungarian Academy of Sciences.
We also acknowledge KIF\"U for awarding us access to computational resources based in Hungary.


\section*{References}


\providecommand{\latin}[1]{#1}
\makeatletter
\providecommand{\doi}
  {\begingroup\let\do\@makeother\dospecials
  \catcode`\{=1 \catcode`\}=2 \doi@aux}
\providecommand{\doi@aux}[1]{\endgroup\texttt{#1}}
\makeatother
\providecommand*\mcitethebibliography{\thebibliography}
\csname @ifundefined\endcsname{endmcitethebibliography}
  {\let\endmcitethebibliography\endthebibliography}{}




\end{document}